# Performance Considerations
# For an Embedded Implementation of OMA DRM 2

Daniel Thull, Roberto Sannino

`daniel.thull@st.com, roberto.sannino@st.com`

STMicroelectronics, Advanced System Technology

Via Olivetti 2, 20041 Agrate Brianza, Italy

## Abstract

*As digital content services gain importance in the mobile world, Digital Rights Management (DRM) applications will become a key component of mobile terminals. This paper examines the effect dedicated hardware macros for specific cryptographic functions have on the performance of a mobile terminal that supports version 2 of the open standard for Digital Rights Management defined by the Open Mobile Alliance (OMA). Following a general description of the standard, the paper contains a detailed analysis of the cryptographic operations that have to be carried out before protected content can be accessed. The combination of this analysis with data on execution times for specific algorithms realized in hardware and software has made it possible to build a model which has allowed us to assert that hardware acceleration for specific cryptographic algorithms can significantly reduce the impact DRM has on a mobile terminal's processing performance and battery life.*

**Keywords**: DRM, Security, Mobile Terminal, Cryptography

## 1. Introduction

The Open Mobile Alliance, a standardization organization for service enablers in the mobile domain with over 350 member companies [1], has recently completed work on version 2 of its open standard for DRM on mobile terminals. With respect to the first version, OMA DRM 2 provides additional features and a significantly higher level of security so as to protect high-value digital content like polyphonic ring-tones, mp3 audio files or video clips on mobile terminals.

Security in OMA DRM 2 is based on a Public Key Infrastructure (PKI) for key distribution and symmetric encryption algorithms for content protection. This paper starts out by describing the actors that are defined by OMA DRM 2 and how they interact in order to grant the final user access to protected content. In chapter 2.4, we take a closer look at the cryptographic operations that are involved in this process. Chapter 0 combines these steps with execution times for different cryptographic algorithms realized in software as well as in hardware and evaluates the impact of hardware acceleration on a mobile terminal's overall performance with respect to execution time and power consumption.

## 2. The Standard

The OMA DRM 2 standard consists of three documents. The DRM specification document [2] defines the communication protocol ROAP as well as general system aspects. The content format for protected media files (Digital Content Format DCF) and the Rights Expression Language (REL), which describes permissions and constraints to govern usage of protected content, are defined in two separate documents.

### 2.1 Actors

The DRM specification document defines four actors that interact with each other in order to provide access to protected digital content to the end-user (see Figure 1). In a procedure not covered by the standard, the Content Issuer (CI), as the owner of digital content, negotiates licenses that grant access to its content with one or more Rights Issuers (RI). Before selling a license to the end-user, the RI sets up a trusted relationship with the DRM Agent, a trusted logical entity residing in the user's terminal. Trust in OMA DRM 2 is based on PKI-certificates issued by a Certification Authority (CA). A valid certificate guarantees that its subject (either the RI or the DRM Agent) adheres to the CA's compliance and robustness rules and can thus be considered trustworthy and secure. Although crucial to the system, the certification process including the definition of compliance and robustness rules is outside the scope of OMA DRM and is left to the business community. The first CA for OMA DRM is called Content Management License Administrator and has been founded in February 2004 [4].

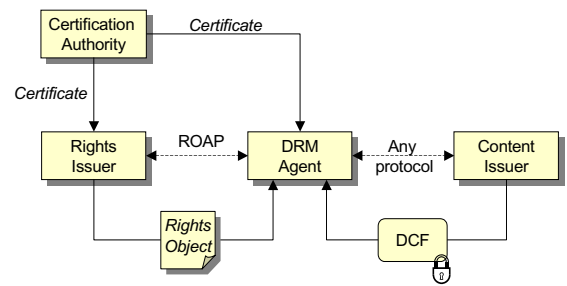

**Figure 1 - OMA DRM 2 Actors.**

## 2.2 Objects

Content and license are delivered to the DRM Agent in two separate logical entities: Content Object (aka DCF although the acronym describes the file format rather than the file itself) and Rights Object (RO). The DCF contains one or more containers that comprise encrypted digital content alongside descriptive meta-data such as author, title and a URL the user may visit in order to obtain a license that allows her to unlock the content.

The Rights Object is realized as an XML file that describes permissions and constraints granted to the DRM Agent when accessing a specific DCF. It also contains the Content Encryption Key ($K_{CEK}$) needed to decrypt the DCF as well as the Rights Encryption Key ($K_{REK}$) with which the former is encrypted. This two-layer symmetric encryption provides a cryptographic way to decouple content and rights and allows building different licenses for the same content without re-encrypting it. $K_{REK}$ itself is encrypted using the DRM Agent's public key, establishing thus a cryptographic chain that can only be dissolved by the holder of the DRM Agent's private key (Figure 2).

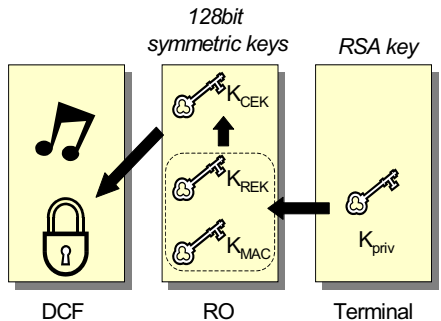

Figure 2 - The cryptographic chain that protects content (no-domain case).

## 2.3 Domains

One important feature of OMA DRM 2 is the possibility to share a license for protected content among a group of devices. In order to do so, the user has to set up a domain and register each participating device with the Rights Issuer, possibly indirectly by using another device as a proxy. During the registration process, the RI relies on a PKI mechanism to provide each trusted device with a secret (symmetric) domain key. This key can subsequently be used by each member device to decrypt $K_{REK}$ of any Domain Rights Object that has been acquired by any member of the group.

By offering the possibility to consume OMA DRM protected content also on devices that cannot directly connect to the RI (the so-called "Unconnected Devices" like mobile mp3 players), the OMA DRM 2 standard broadens its scope well beyond mobile phones. Any device that adheres to a Certification Authority's compliance and robustness rules and owns a valid certificate can thus be used to access protected content.

## 2.4 Phases

The consumption process of DRM protected media can be divided in four phases: Registration, Acquisition, Installation, and Consumption.

### 2.4.1 Registration – Establishing Trust

In order to prevent leakage of clear content from compromised devices, the Rights Issuer delivers Rights Objects only to trusted DRM Agents, ie those whose certificate has not expired or been revoked. In OMA DRM 2 a trusted relationship gets established when a DRM Agent registers with an RI by following the 4-pass Rights Object Acquisition Protocol (ROAP).

During the first phase of the registration process, both partners advertise their capabilities to each other. This may result in an agreement to use any cryptographic algorithm other than the ones mandated by the standard (see chapter 2.4.5). In the second registration-step, the DRM Agent sends its PKI certificate in a digitally signed message (RegistrationRequest) to the RI. After having verified the message signature and the Agent-certificate's validity – possibly using an Online Certificate Status Protocol (OCSP) request – the RI responds by sending the RegistrationResponse message to the DRM Agent. This message contains the RI's certificate as well as a valid OCSP response for it, indicating whether the certificate has been revoked [3].

Upon reception, the Agent verifies the message signature and the validity of the RI's certificate as well as the OCSP response. If no check has failed, the DRM Agent saves information on the relationship with this specific RI in the RI Context. This data object represents the trusted relationship from the DRM Agent's point of view and its existence, integrity and validity must be verified prior to any future interaction with the RI, such as RO Acquisition.

### 2.4.2 Acquisition – The Rights Object

In order to acquire a license for a DCF, the DRM Agent checks the existence and validity of an RI Context and sends a digitally signed RORequest message specifying the desired license (Rights Object ID) to the Rights Issuer. If a trusted relationship exists between the parties and payment has been taken care of (the payment process is not within scope of OMA DRM), the RI responds with a digitally signed ROResponse message which contains the protected Rights Object.
The Rights Object is integrity-protected by a Message Authentication Code (MAC) and contains a list of Content Object IDs and their respective usage permissions. The MAC-key $K_{MAC}$ is protected together with $K_{REK}$, using a PKI mechanism. The RO is thus not only integrity but also authenticity protected.

### 2.4.3 Installation – Unwrapping the Keys

After the DRM Agent has extracted the Rights Object from the ROResponse message, it must control its integrity and authenticity before installing the RO on the device. In order to do so, the Agent decrypts $C_1$ (the first 1024 bits of C which is contained in the RO – see Figure 3) using its private key and obtains Z. Applying the key derivation function KDF to Z yields KEK (key encryption key) which is then used to decrypt $C_2$, the last 256 bits of C. As a result of this last decryption, the DRM Agent obtains a concatenation of $K_{MAC}$ and $K_{REK}$ in clear.

After it has successfully checked RO integrity and authenticity using $K_{MAC}$, the DRM Agent must verify the Rights Object's signature in case it is present, using the RI's public key. This signature is made over certain parts of the Rights Object and is mandatory only for Domain ROs. It is however possible to also sign Device ROs.

The OMA DRM 2 standard does not define technical details on how the DRM Agent shall store Rights Objects and DCFs. This is left to the Certification Authorities to define in their robustness rules. At the time of writing this article, the Content Management License Administrator (CMLA) is the only Certification Authority (CA) for OMA DRM 2 [4]. Although different CAs are likely to issue different robustness rules, an obvious requirement that should be common to all is that content and rights are stored in a secure manner. In order for this to happen, four things have to be ensured:

- **Content confidentiality** – Since secure memory is an extremely scarce and costly resource in a mobile terminal, DCFs do not get stored in clear. Thus, content confidentiality is guaranteed.
- **RO integrity** is ensured by the MAC that is included in the Rights Object. This implicitly also ensures the connection between RO and DCF since a hash value of the DCF is included in the Rights Object.
- **RO authenticity** – The authenticity of the received Rights Object has been verified when $K_{REK}$ was decrypted successfully. Since it is assumed that only trusted DRM Agents can successfully complete this operation, there is no need to further protect and check authenticity.
- **$K_{CEK}$, $K_{REK}$ and $K_{MAC}$ confidentiality** – Also after installation, $K_{CEK}$ gets protected by $K_{REK}$. This makes sense because there might be more than one Rights Object for a DCF, so the DRM Agent would have to keep information on the associated keys anyway. In the original Rights Object, $K_{REK}$ and $K_{MAC}$ get protected by a public key encryption. Since PKI algorithms are very performance intensive, it is desirable to replace them with simpler ones where possible. In this case we chose to substitute the PKI-encryption with a symmetric encryption using a device-generated key $K_{DEV}$ when installing the RO. This is possible because the RO will only be consumed by the installing DRM Agent. This means that the PKI algorithm's main purpose (ie to allow two strangers to share a secret over an insecure channel) is no longer needed and it can be substituted by a less calculation-intensive symmetric cipher. Encrypting $K_{REK}$ and $K_{MAC}$ with $K_{DEV}$ yields $C_{2dev}$ that can be stored safely in any type of memory (see Figure 3).

### 2.4.4 Consumption – Steps to Follow for Every Access

Every time the user wants to access protected content, the DRM Agent has to perform the following cryptographic processing steps:
1. Decrypt $C_{2dev}$ using $K_{DEV}$
2. Verify RO integrity by checking its MAC
3. Verify DCF integrity by calculating its Hash value and comparing it to the one from within the RO.

This is important to bear in mind since these processing steps also apply to small files like ring-tones, because they cannot be stored in clear since secure memory is extremely costly in mobile terminals.

### 2.4.5 Standard Cryptographic Algorithms

So far we have described the cryptographic operations required by OMA DRM 2 in generic terms. This has been done deliberately since the standard provides the possibility to use other algorithms than the ones that are pre-defined. For the following considerations however, we relied on the standard algorithms, in particular:

- SHA-1 as Hash function.
- HMAC SHA-1 as MAC algorithm.
- 128-bit AES WRAP for en-/decrypting keys.
- 128-bit AES CBC for content en-/decryption.
- RSA-PSSA as signature scheme (using the RSASP1 and RSAVP1 primitives as defined in [5]).
- KDF2 as key derivation function as described in [2].
- 1024-bit RSA as PKI function (using the RSAEP and RSADP primitives as defined in [5]).

A detailed study of the OMA DRM 2 specifications allowed us to build a Java software model of the standard including Rights Issuer, Content Issuer and DRM Agent. The deeper understanding of the cryptographic system implications we obtained from this work, resulted in information about eg the ROAP message file sizes and yielded a list of cryptographic operations carried out in each of the four phases that have been identified above.

For the sake of simplicity we have made close approximations when compiling this list wherever too many details would not have augmented clarity. One example for this is the EMSA-PSS message-encoding mechanism described in [5], which we have approximated with just one hash function over the message code.

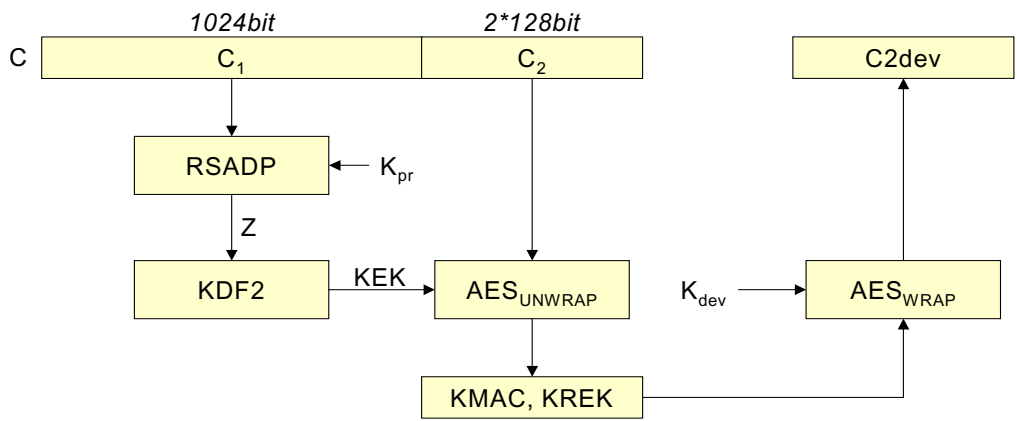

**Figure 3 - Extraction, decryption and encryption of $K_{REK}$ and $K_{MAC}$ during the installation process.**

| Algorithm | Software [cycles] | Hardware [cycles] |
|---|---|---|
| AES Encryption | 360 + 830/128 bit | 10/128 bit |
| AES Decryption | 950 + 830/128 bit | 10 + 10/128 bit |
| SHA-1 | 400/128 bit | 20/128 bit |
| HMAC SHA-1 | 1200 + 400/128 bit | 240 + 20/128 bit |
| RSA 1024 Public Key Op | 2,160,000/1024 bit | 10,000/1024 bit |
| RSA 1024 Private Key Op | 3,774,0000/1024 bit | 260,000/1024 bit |

**Table 1 - Execution times for different cryptographic algorithms in hardware and software.**

## 3. Costs – Time and Energy

The impact of DRM on overall terminal performance depends to a great extent on the system architecture that provides the supporting functionality. From an end user's perspective, the most important performance-dimensions of a mobile terminal are price, processing time and energy consumption (ie, battery lifetime). A system architect must find he optimal tradeoff between these factors when deciding on whether to support functionality in hardware or in software [9]. This paragraph deliberately neglects DRM impact on monetary terminal costs and concentrates on processing-performance and energy consumption as a result of different architecture-choices.

The underlying assumption for the following considerations is that a mobile terminal contains a System-on-Chip (SoC) that provides all application-related functionality. This element is also known as Application Processor. The SoC consists of various dedicated hardware modules, a general purpose processor core and secure on-chip memory. All these elements are connected by a system bus.

For our performance considerations we relied on the system know-how we had acquired while implementing OMA DRM 2 in Java on a PC as well as publicly available performance figures for the necessary cryptographic algorithms in software and hardware (see Table 1). Following this system-level approach, we are currently conducting more detailed experiments that allow for a more accurate consideration of energy consumption. For this paper we assumed energy consumption to be directly related to processing performance. Hence, the estimation figures we obtained regarding processing time can be taken as a first very rough estimate of the effect, the DRM application has on a terminal's energy consumption, although the inherent inefficiencies of protocol-overhead and other non-cryptographic functionality have not been considered.

It is important to note that we deliberately neglected system-related time consumption such as cache-misses or bus-conflicts in our considerations as these depend on factors such as concurrent applications running that are not strictly related to OMA DRM and there is no generic model that would allow simulating their impact. Furthermore, we concentrated on cryptography-related aspects of OMA DRM and did not take the overhead caused by protocol processing technology like XML parsing into consideration when evaluating the impact of DRM since these components cannot easily be accelerated by dedicated hardware cells.

When facing the challenge to implement a DRM Agent on a mobile terminal, a system designer has to identify crucial processing intensive parts of the application and decide whether to provide these using dedicated hardware cells within a SoC or rather software running on a general purpose processor.

In OMA DRM, the most processing intensive operations that can be realized in hardware are cryptography-related. Security- and price-related considerations apart, dedicated cryptographic hardware modules offer two benefits as compared to software running on a general purpose processor: they are much faster and leave the processor free to do other jobs in parallel.

In order to determine the benefit of different cryptographic hardware accelerators to the OMA DRM application, we calculated the overall processing time needed to perform certain standard operations such as acquiring a Rights Object as well as the relative time spent for each algorithm.

shows the processing time in clock-cycles per data block for each cryptographic algorithm in software (tested on the ARM9 processor) and hardware (with a clock-frequency of less than 200 MHz). Execution times for AES and SHA-1 in hardware were obtained from [6] whereas the software figures were obtained from internal experiments using standard implementations. Numbers regarding RSA were taken from [7] in the hardware and from [8] in the software case. The constant offset-values for AES and HMAC are due to key-scheduling (AES) and hashing on fixed-length data (HMAC).

We evaluated three architecture variants: a purely software-based approach, a mixed case in which AES and SHA-1 (and thus also HMAC SHA-1) are provided by hardware modules and RSA by software as well as a pure hardware case with dedicated modules for each algorithm. Clock-frequency was assumed to be 200 MHz in each case.

## 4. Results – Evaluation of Two Use Cases

Since it is the final user who in the end evaluates terminal performance, measuring DRM's impact cannot be done by examining isolated cryptographic operations but rather by starting from typical use cases. We have identified two of them, decomposed them into single cryptographic steps and based our calculations on the resulting model. The two use cases are:

- **Music Player:** the user has access to an encrypted content file (DCF) of 3.5 Mbytes. In order to obtain a license, she registers with an RI, acquires the license and installs it. She then listens to the track five times.
- **Ringtone:** the user has downloaded an encrypted high-quality polyphonic ringtone file (size 30 Kbytes). She registers with the RI, obtains a license and installs it. Every time her phone rings, the DRM Agent must now access the encrypted file that is governed by the usage rights in the Rights Object. We assume that she receives 25 calls.

The two use cases differ mainly in the size of the encrypted file and in the number of playbacks. For the sake of simplicity, domain functionality has not been included in the examples.

Figure 5 illustrates the percentage of total processing time the processor spends for each cryptographic algorithm (realized as a software program) in both use cases. Because of the larger file size, AES and SHA-1 become much more important in the Music Player use case whereas in the Ringtone use case the PKI algorithms that prevail during the registration-/installation-phases play a greater role. The effect of dedicated hardware macros for AES and SHA-1 has thus a much greater effect in the Music Player use case as can be seen in Figure 6 and Figure 7.

In the music player use case, total processing time can be cut to almost a tenth of the value obtained from a pure software implementation by realizing AES and SHA-1 as dedicated hardware macros. In the Ringtone use case, the significant step occurs when providing PKI hardware support.

Hardware acceleration for PKI algorithms has only limited benefits in both use cases from a performance point of view. Since PKI algorithms get only used in the initiating application phases and their execution time does not depend on the DCF size, the absolute figures are identical for both use cases. Given that they total to roughly 600ms (the second column in Figure 7), it is arguable whether or not the costs of a dedicated hardware cell (in terms of transistor gates) are justified by these use cases of the OMA DRM application.

## 5. Conclusions and Future Work

As digital content services gain importance in the mobile world, DRM applications will become a key component of mobile terminals. Given the broad support among terminal manufacturers and it being an open standard, OMA DRM 2 is bound to assume a strong position in the standard battle. A detailed study of the cryptographic operations required by OMA DRM 2 has shown that the impact DRM has on processing performance and battery life can be significantly reduced by incorporating hardware acceleration for specific algorithms. For a complete evaluation of the hardware/software partitioning however, also security related aspects have to be considered.

We are currently conducting more detailed simulations regarding energy consumption of dedicated hardware macros for the cryptographic algorithms presented in this article. First results seem to indicate that the gap between software and hardware realizations in this case is even wider than for processing time.

## 6. Acknowledgments

We would like to thank Guido Bertoni and Roberto Zafalon for their support in writing this article.

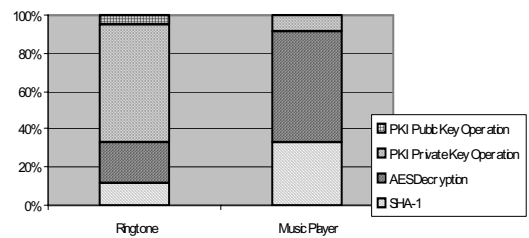

**Figure 5 – Relative importance of cryptographic algorithms in both use cases.**

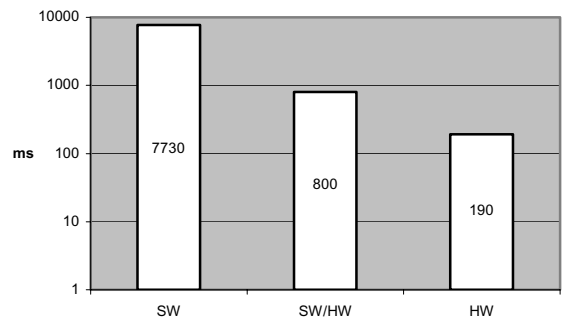

**Figure 6 – Execution times for three implementation variants in the Music Player use case.**

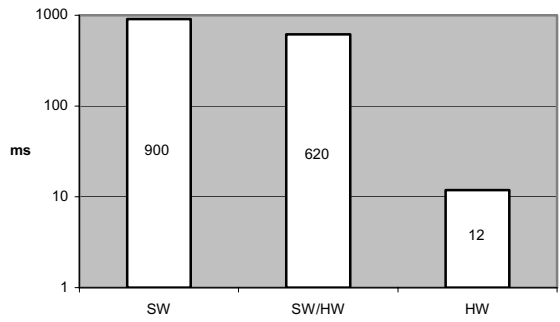

**Figure 7 – Execution times for three architecture variants in the Ringtone use case.**